# The Sub-Eddington Boundary for the Quasar Mass–Luminosity Plane: A Theoretical Perspective

**David Garofalo** [1,*], **Damian J. Christian** [2] **and Andrew M. Jones** [3]

1. Department of Physics, Kennesaw State University, Marietta 30060, GA, USA
2. Department of Physics & Astronomy, California State University, Northridge 91330, CA, USA; damian.christian@csun.edu
3. Department of Mathematics, Boise State University, Boise 83725, ID, USA; andrewjones237@u.boisestate.edu
* Correspondence: dgarofal@kennesaw.edu

**Abstract:** By exploring more than sixty thousand quasars from the Sloan Digital Sky Survey Data Release 5, Steinhardt & Elvis discovered a sub-Eddington boundary and a redshift-dependent drop-off at higher black hole mass, possible clues to the growth history of massive black holes. Our contribution to this special issue of Universe amounts to an application of a model for black hole accretion and jet formation to these observations. For illustrative purposes, we include ~100,000 data points from the Sloan Digital Sky Survey Data Release 7 where the sub-Eddington boundary is also visible and propose a theoretical picture that explains these features. By appealing to thin disk theory and both the lower accretion efficiency and the time evolution of jetted quasars compared to non-jetted quasars in our "gap paradigm", we explain two features of the sub-Eddington boundary. First, we show that a drop-off on the quasar mass-luminosity plane for larger black hole mass occurs at all redshifts. But the fraction of jetted quasars is directly related to the merger function in this paradigm, which means the jetted quasar fraction drops with decrease in redshift, which allows us to explain a second feature of the sub-Eddington boundary, namely a redshift dependence of the slope of the quasar mass–luminosity boundary at high black hole mass stemming from a change in radiative efficiency with time. We are able to reproduce the mass dependence of, as well as the oscillating behavior in, the slope of the sub-Eddington boundary as a function of time. The basic physical idea involves retrograde accretion occurring only for a subset of the more massive black holes, which implies that most spinning black holes in our model are prograde accretors. In short, this paper amounts to a qualitative overview of how a sub-Eddington boundary naturally emerges in the gap paradigm.

## 1. Introduction

Active Galactic Nuclei (AGN) have been found to obey their Eddington luminosity limit, and even supermassive black holes in quiescent galaxies can be actively accreting at rates of ≈0.1% that of Eddington (Peterson 2014). Building on the work of [1] showing that black holes grow predominantly by accreting close to their Eddington limit, [2] explored the quasar mass-luminosity plane for 62,185 quasars from the Sloan Digital Sky Survey (SDSS) and discovered an additional surprising and as yet unexplained 'Sub-Eddington Boundary' (see also [1] Figure 6 including objects at higher redshift where the sub-Eddington boundary appears). Their results are shown in Figure 1 (taken from Figure 1 of [2]), showing that quasars tend to not only obey the Eddington limit, but seem to shy away from it in a mass-dependent way such that at higher black hole mass there is a "drop-off", with the slope of the boundary line decreasing. In addition, they also find the drop-off to increase further beyond some redshift threshold, in the sense that its slope is smaller. We find similar behavior for a "drop-off" in luminosity for higher mass black holes and a sub-Eddington boundary in our analysis of the SDSS DR7 data [3] shown in our Figure 2. Understanding this behavior promises to shed light on the history of black hole growth. The existence of a sub-Eddington boundary, however, is quite controversial [4–6].

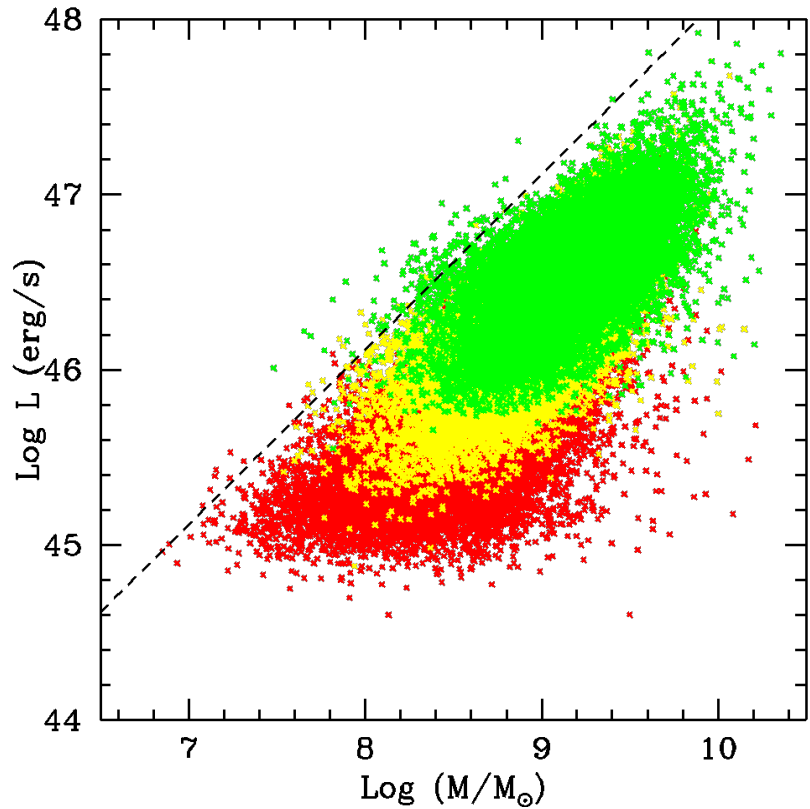

**Figure 1.** The quasar mass-luminosity plane from Figure 1 of [2]. The dashed line represents the Eddington boundary. Red indicates a redshift value of 0.2 < z < 0.8, yellow indicates 0.8 < z < 1.4 and green is for 1.4 < z < 2. Black hole mass is estimated from broad line region emission lines in combination with the virial velocity (details in [2]).

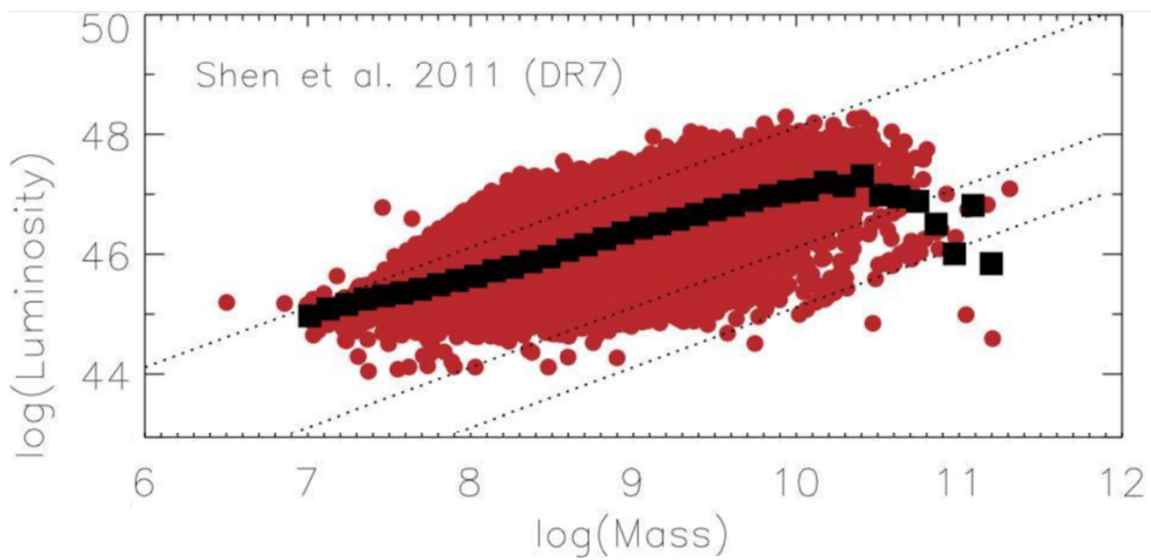

**Figure 2.** The plot of data from Sloan Digital Sky Survey (SDSS) Data Release 7 including average luminosity as a function of mass ([3])—the top line is the Eddington boundary, and two additional lines at 1% and 0.1% of Eddington. Units are the same as in Figure 1. Note the absence of objects

reaching or even approaching the Eddington limit at log M = 10.5 (i.e., just above Log L = 48) and beyond. Although the possible missing low luminosity points could change the slope of the black line, it would not change the fact that whereas some of the data for small values of log M exceed the Eddington boundary, at high log M no such data exceed it. In fact, the data at high log M actually drop off sharply from the Eddington boundary. Note how the number of points above the Eddington limit is greater here compared to SDSS 5 of Figure 1. Of course, SDSS 7 almost doubles the number of objects.

Whether or not a sub-Eddington boundary exists, we show how it may emerge from our theoretical framework. We reproduce these features of the sub-Eddington boundary by adopting the theoretical framework known as the 'gap paradigm' for black hole accretion and jet formation ([7]), according to which retrograde black hole accretion served an important role in the high redshift Universe. Our goal is to describe a picture in which the sub-Eddington boundary emerges from a prescription for redshift z~1.5–2 quasar evolution focused on retrograde accreting black holes, which is both difficult to explore (e.g., [8]) and where a variety of processes appear to be at work (e.g., [9]), by applying the same ideas that allowed the model to shed light on the high redshift AGN phenomenon ([10,11]). In order to accomplish our goal, we appeal to the stabilizing effect of larger black hole mass on retrograde accretion in tandem with its implication for the redshift distribution of different types of active galactic nuclei ([11–14]), as we now explain.

When the black hole is massive enough compared to the mass content of the accreting material, the post-merger cold gas funneled toward the supermassive black hole will eventually settle into a thin disk configuration that is equally likely to form in either prograde or retrograde configuration. But when the relative mass constraint between a black hole and accretion disk is not satisfied, retrograde accretion configurations are absent. This constraint comes to us from [12] which we will describe quantitatively. The essence of that work, however, is that the most massive black holes have equal probability of being surrounded by either a retrograde or prograde accretion disk while that probability for retrograde configurations drops as the black hole mass decreases. We will provide quantitative prescriptions as a function of redshift for the fraction of retrograde forming accreting black holes in post mergers that are informed by models of black hole spin formation. Because these accreting black holes are assumed to result from mergers, our mass constraint is modulated by the merger function, which evolves with time. We will use a merger function that is a combination of observations and simulations that ends up producing an almost linear decrease in the merger rate as the redshift decreases from about z = 2. As the merger rate drops, the fraction of accreting black holes that are the product of mergers decreases compared to the entire AGN population, which eventually becomes dominated or fed by secular processes. Consequently, the fraction of retrograde accreting black holes drops and this has an effect on the average efficiency of the Eddington-limited accreting black holes. In short, there is both a black hole mass as well as a redshift dependence on the fraction of retrograde accreting black holes which we will make quantitative. In making clear the spirit of our work, we emphasize that our explanation is contingent or dependent on a simple prescription concerning the nature of the black hole engine. Our goal is to show how much this simple picture can account for. The simplicity of the model, however, does not imply that the physics of the greater galaxy is irrelevant to the overall character of the quasar, but simply that the black hole engine leaves an indelible imprint. We will construct a mass–luminosity plane from theory and show how the aforementioned features of the sub-Eddington boundary emerge. In Section 2 we construct a mass–luminosity plane by averaging over a range of luminosities for fixed values of black hole mass to show how the slope at higher mass drops to lower values. We then explore how the value of this slope depends on redshift. In Section 3, we conclude.

**2. Analysis and Discussion**

*2.1. IIa-Slope Vs. Mass of the Sub-Eddington Boundary*

From Figures 1 and 2, we recognize how accreting black holes are not restricted to radiate at their Eddington values, but span a range of about 1 or 2 orders of magnitude for the available data.

More relevant to our work is the fact that the data also suggests that the Eddington limit tends to be obeyed, the degree to which that limit is obeyed being the subject of our paper. Surprisingly, the way the AGN population obeys the Eddington limit does not appear to be scale-invariant since Figures 1 and 2 indicate that smaller mass black holes seem more willing to approach or violate that limit compared to their high mass counterparts, albeit by a small amount. We see that although a variety of luminosities are possible for any black hole mass, at higher black hole mass, the Eddington limit becomes an increasingly significant constraint, with fewer objects approaching or violating that limit. There seems to be an apparent scale-invariant violation in the way the luminosity range clusters about the Eddington limit. We will argue that this is a feature of the difference in the average type of accretion for lower versus higher black hole mass, with differences due to different accretion efficiencies, resulting directly from the above-mentioned accretion disk stability dependence on black hole mass (i.e., on the fraction of retrograde accreting black holes).

We now highlight the elements of standard thin disk accretion that are needed to understand our explanation of the sub-Eddington boundary. While low angular momentum advection-dominated and super Eddington accretion flows have significantly lower efficiencies [15,16]), flows which radiate closest to the Eddington limit are standard thin disk accretors. Because we are interested in exploring the behavior of accretion close to the Eddington boundary so as to explain the sub-Eddington boundary, we will ignore less efficient accretion models for most of the discussion and focus on standard thin disk, [17] accretion, or more precisely, on the relativistic version of [18] The reason is that less efficient accretion fails by definition to reach the Eddington limit. However, in order to explain the observations of [2] at redshift less than about 0.5, we will need to look at the time evolution of massive accreting black holes, which will force us to deal with low efficiency, ADAF accretion. For our initial exploration, however, we focus only on thin relativistic accretion disks, which are characterized by stable circular orbits that are truncated at the innermost stable circular orbit (ISCO), a value that depends on black hole spin as well as on the orientation of the accretion flow. The non-relativistic version of the dissipation as a function of radial distance in the disk depends on the location of the inner boundary $r_{isco}$ according to:

$$D(r) = \frac{3GM}{4\pi r^3}\frac{dM}{dt}[1 - (\frac{r_{isco}}{r})^{1/2}] \qquad (1)$$

with G Newton's constant, M the black hole mass, and dM/dt the accretion rate. When integrated over the two faces of the thin disk, one obtains the luminosity as:

$$L_{disk} = \frac{GM}{2r_{isco}}\frac{dM}{dt} \qquad (2)$$

The disk luminosity is, therefore, ISCO dependent. The smaller the ISCO is, the larger the total luminosity. Because (as emphasized in the Introduction) the larger mass (M) objects less strongly constrain disk orientation and are thus progressively more likely in the paradigm to be surrounded by retrograde or prograde accretion, the larger M objects are also more likely to have ISCO values that span a wider range than those with smaller M values. The ISCO values for the larger M objects, therefore, tend to span much more of the range 1.23 $GM/c^2$–9 $GM/c^2$ (with c the speed of light constant) while the smaller M objects tend to span the smaller range 1.23 $GM/c^2$–6 $GM/c^2$. The former range is associated with ISCO values spanning the full retrograde through prograde accretion from high to low black hole spin. The latter range, on the other hand, is associated with prograde-only accretion from high to low black hole spin, which we employ to model lower mass accreting black holes (see, e.g., the AGN branching-tree diagram in [11]. In short, associated with the larger fraction of retrograde accreting black holes at larger black hole mass, there is also a greater range of ISCO values. And as Equation (2) illustrates, ISCO values determine the luminous efficiency. We show the relationship between the ISCO value and the disk efficiency in Table 1 with the ISCO value in gravitational radii and plot them in Figure 3.

Table 1. The left column shows the innermost stable circular orbit (ISCO) values in gravitational radii, while the right column reports the disk efficiency normalized to the highest prograde spin case.

| ISCO ($r_g$) | H |
|:---:|:---:|
| 9.0 | 0.111 |
| 8.7 | 0.115 |
| 8.5 | 0.118 |
| 8.2 | 0.122 |
| 7.9 | 0.127 |
| 7.5 | 0.133 |
| 7.2 | 0.139 |
| 6.9 | 0.145 |
| 6.6 | 0.152 |
| 6.3 | 0.159 |
| 6.0 | 0.167 |
| 5.8 | 0.172 |
| 5.33 | 0.188 |
| 5.0 | 0.200 |
| 4.61 | 0.217 |
| 4.3 | 0.233 |
| 3.83 | 0.261 |
| 3.3 | 0.303 |
| 2.9 | 0.345 |
| 2.2 | 0.455 |
| 1.8 | 0.555 |
| 1.5 | 0.667 |
| 1.237 | 0.808 |
| 1.00 | 1.00 |

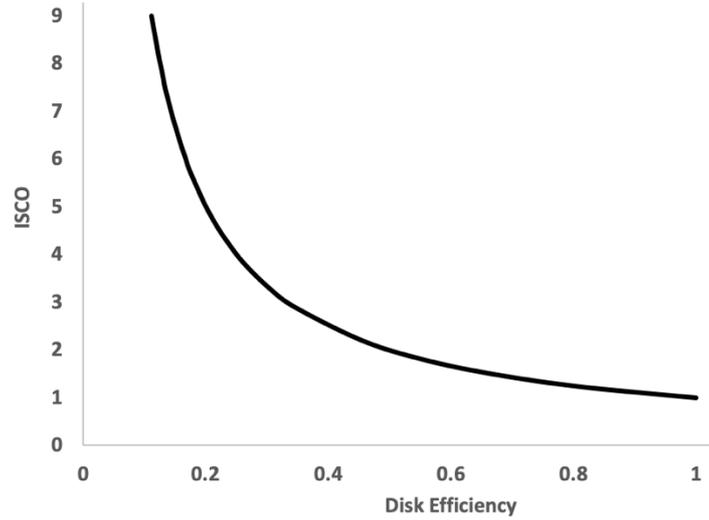

**Figure 3.** ISCO values and disk efficiency from Table 1.

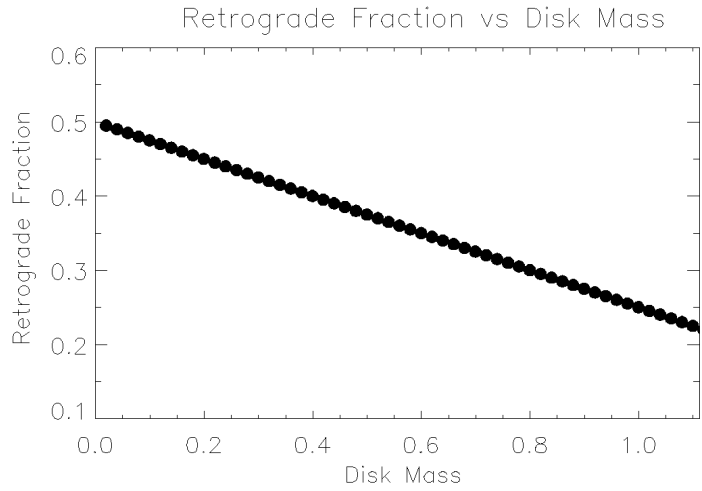

**Figure 4.** Retrograde fraction as a function of disk mass in terms of the black hole mass such that a value of 1 indicates equality between disk and black hole from the analysis of [12] via our Equation (3). As the total mass of the disk decreases compared to the black hole, its total angular momentum decreases as well, and the retrograde fraction approaches 0.5, reflecting the random nature of the disk configuration around the rotating black hole.

Since we are interested in scale-invariant explanations, we impose scale-invariance in the accretion rate, i.e., we assume a fixed average Eddington accretion rate across the black hole mass population. This is not unusual, nor restrictive, but a natural consequence of the fact that galactic processes that funnel gas toward the black hole are decoupled from the black hole sphere of influence (e.g., [19–22]). This leaves us with one free parameter, the ISCO value. Because of the absence of any physical constraints, we allow black hole spin to vary randomly over a large range in spin except at the very high end of the spin where simulations suggest the highest spins are more difficult to obtain in mergers ([23]). This means that most values of black hole spin are weighted equally. This does not mean, however, that prograde, as well as retrograde systems, are equally likely and this is a crucial distinction that imposes non-negligible constraints on the results. As discussed above, the retrograde fraction has both a mass as well as a redshift dependence. With that in mind, and with an eye toward constructing a theoretical version of Figure 1, we proceed to obtain an average luminosity for a given black hole mass. We average over all possible accretion disks that differ in ISCO values to determine an average disk luminosity as a function of black hole mass by imposing the following prescription for the retrograde accreting fraction. In attempting to impose the physical constraints from King et al. (2005) as discussed further below, we assume that no retrograde accreting black holes form at M < $10^7$ $M_\odot$, where $M_\odot$ is one solar mass. A more sophisticated prescription for retrograde formation at lower black hole mass would not qualitatively change any of our conclusions. This means that in the aftermath of mergers that funnel cold gas to the central merging black holes, an accretion disk forms whose angular momentum vector is aligned with that of the black hole, which constitutes prograde accretion. As discussed in the Introduction, the black hole is not massive enough to remain stable in a retrograde configuration so any retrograde forming disks simply flip to a prograde state. As we consider more massive black holes on the order of $10^8$ $M_\odot$ or more, however, we relax our constraint by allowing the fraction of retrograde objects to increase progressively toward 50% at high masses near $10^9$ $M_\odot$, implying that both prograde and retrograde disks become equally stable for large black holes. Our prescription for the retrograde fraction is grounded in the work of [12] who showed that the relative total angular momentum of the accretion disk to that of the black hole determines the retrograde fraction according to:

$$f = \tfrac{1}{2}(1 - J_d/2J_h) \qquad (3)$$

where $J_d$ and $J_h$ are the total angular momentum of the disk and black hole, respectively. Because both angular momenta are obtained as an integral that is linearly dependent on mass, the fraction of retrograde occurrences can be translated into a mass ratio, with a higher fraction corresponding to lower accretion mass. Smaller total accreting mass compared to the black hole mass results in stability in either prograde or retrograde cases and the fraction tends to ½ (Figure 4). Whereas the range of accretion rates is assumed scale-invariant, the ratio $J_d/J_h$ is not. This comes from the fact that the amount of cold gas that makes it into the central region to form the accretion disk is decoupled from black hole gravity. For mergers of larger galaxies, the scale invariance in the range of accretion means there is a larger range in that rate. Black hole gravity cannot regulate or determine the amount of cold gas that is funneled into its accretion disk because black hole feedback has yet to occur. Bigger black holes live in bigger galaxies and their mergers will produce a larger range of accretion rates that reach higher values that scale with the galaxy size and thus black hole mass. But because the formation of the disk is uncoupled from the black hole, that larger range means that accretion disks with a wider range of total mass and thus total angular momentum are possible, which means the existence of an increasing subset (compared to smaller black holes) of the population with small(er) total disk angular momentum compared to that of the black hole. This means that on average for galaxies with larger black holes, the angular momentum fraction $J_d/J_h$ tends to be smaller. Scale invariance does not ensure that larger black holes become surrounded by comparatively larger disks with larger total angular momentum. It, instead, makes it possible for larger black holes to be fed by a wider range of mass for the accretion disk, some of which have small mass compared to the black hole. This is true at the formation of the disk as well as at later times. In fact, the ratio of angular momenta becomes progressively skewed toward the black hole once accretion settles into a steady state since accretion grows the black hole mass and its total angular momentum but does not grow the disk mass or its angular momentum.

Equation (3) is not the end of the story, however. Models of black hole spin formation and evolution suggest that high spinning black holes formed from mergers are a small subset of the total population of accreting black holes ([23,24]). Therefore, on top of the stability issue explored in Equation (3), we need to lower the occurrence of high-spinning retrograde accretion configurations relative to medium and low spin configurations. To capture this, we impose an upper boundary to the spin of 0.9 which limits the disk efficiency in retrograde mode to no less than about 0.115 which corresponds to an ISCO value of 8.7 gravitational radii (see Table 1).

In addition to the mass dependence, the assumption that only mergers can lead to retrograde configurations implies that the retrograde fraction also depends on the behavior of the merger function which is changing with time. Because we are using the merger function in this Section, we now discuss the time evolution of the retrograde fraction that is affected by the merger rate. In Section IIb we will discuss more general time evolution of the retrograde fraction that comes about as a result of both the evolution of the merger rate as well as the evolution of originally retrograde accreting black holes as determined by the accretion process itself. In order to quantitatively deal with the effect on the retrograde fraction as determined by the evolution in time of the merger rate, we use the merger function results of [25] to explain our results for Figure 5. We determine an average merger rate from Figure 6 values for the redshift range 1.5–2 and an average merger rate for the redshift range 0.5–1, obtaining 0.11 and 0.0625 respectively. The merger rate has, therefore, dropped by a factor of 1.76 over this redshift range. Because retrograde accreting thin disks are assumed only to result from mergers, the fraction of possible retrograde accreting thin disks drops to about a factor of 0.6 (the inverse of 1.76) in the redshift range 0.5–1 compared to the redshift range 1.5–2. Therefore, a factor of about 0.6 is the ratio between the retrograde fraction at these two redshift values. However, this is the ratio for black holes at the highest mass, equal to or above $10^{10}$ solar masses, where the retrograde fraction is closest to ½. As the mass of the black hole decreases, so does the retrograde fraction but such that at a threshold mass value, the retrograde fraction becomes zero. Hence, the two curves in Figure 5 must converge. In Figure 5 we show this redshift dependence with the red data representing the fraction of post-merger systems accreting in retrograde mode at higher redshift close to 2 when the merger function is at its peak (Figure 6), while the blue data represents the retrograde fraction at lower redshift when the merger rate has dropped and with it the retrograde fraction. Figures 4, 5, and 6 are combined to produce the time dependence of the retrograde fraction that is needed to determine the evolution of the quasar mass–luminosity function.

Our basic result for the mass dependence is that the average dimensionless ISCO value (i.e., $r_{isco}^{dim}$ = $\frac{r_{isco}}{\frac{GM}{c^2}}$) increases with M due to the greater fraction of retrograde accreting black holes. Because the accretion rate is assumed scale-invariant, the increased fraction of retrograde accreting objects will result in an average luminosity that is less than the Eddington luminosity as the black hole mass increases. The slope of average luminosity versus mass will drop away from the Eddington boundary as the black hole mass increases. Another way of restating this is to recognize that average accretion efficiency drops as the black hole mass increases due to a greater fraction of less efficient retrograde objects. In order to show this, we produce an average luminosity for each black hole mass that reflects this one parameter difference in the accreting black hole population, i.e., an increase in retrograde accreting objects with larger black hole mass. And this prescription (Figure 5) increases with black hole mass for both redshifts as a result of their connection to the merger fraction obtained from [25]— Figure 6. When the post-merger black hole is less than about $10^8$ M$_\odot$, we prescribe a cutoff for simplicity such that 100% of accreting black holes end up surrounded by prograde disks that are associated with larger efficiencies (i.e., smaller average ISCO value). This cutoff will not change our results qualitatively. When the black hole mass is large enough to ensure stability, the fraction of retrograde accreting black holes approaches the value that reflects the randomness of both the angular momentum vector of the black hole and that of the accretion disk, namely 50% ([12]), so both configurations become equally likely.

Our results are displayed in Figure 7. Note the drop-off at higher black hole mass, where the slope above ≈ 4 × $10^8$ M$_\odot$ decreases regardless of redshift. This is a direct consequence of the fact that accreting supermassive black holes span a wider range in efficiency (i.e., a larger range in ISCO values) due to the fact that retrograde orientations have increased occurrence in the high mass accreting black hole population. We have associated the highest prograde spin at 100% efficiency (see Table 1) with the Eddington limit as a normalization condition. If all accreting black holes were chosen to spin at their maximum prograde value, our results for Figure 7 would match the Eddington limit. But since prograde-only black holes are assumed to possess a range of prograde spin values, the average amounts to 0.4 of the Eddington limit. The focus should therefore be on the relative difference in the data in Figure 7 (i.e., on the drop for points associated with black hole masses above a threshold limit), and not on the actual Eddington value of each data point. The exact values for the average luminosity as a function of black hole mass plotted in Figure 7 for the two redshift values are reported in Table 2.

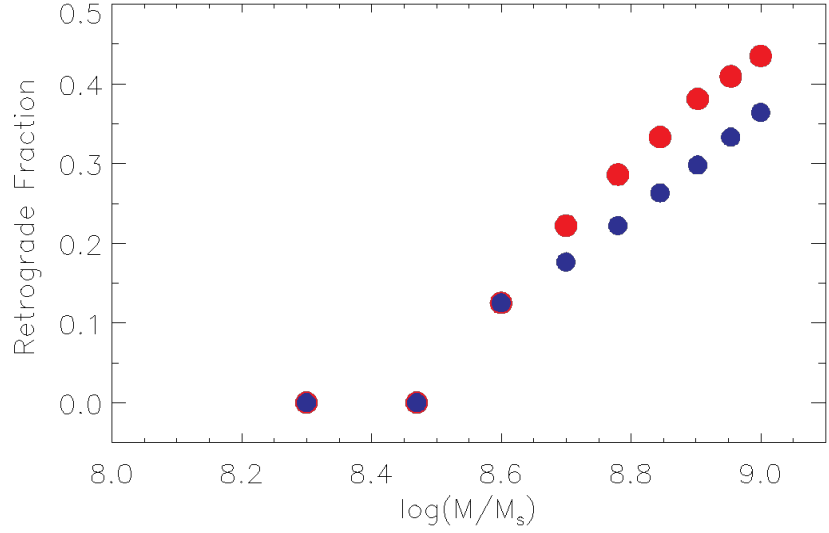

**Figure 5.** Theoretical prescription for the fraction of accreting black holes that form retrograde thin disks as a function of black hole mass. Because accretion around massive black holes is stable, the high mass black holes are progressively more likely to be in either prograde or retrograde configurations, which is why the fraction approaches 0.5 at high mass. The red represents the behavior in the redshift range of 1.5–2 when the merger function is larger (Figure 6) while the blue represents the behavior when the redshift is closer to 1 and the merger function has dropped (Figure 6).

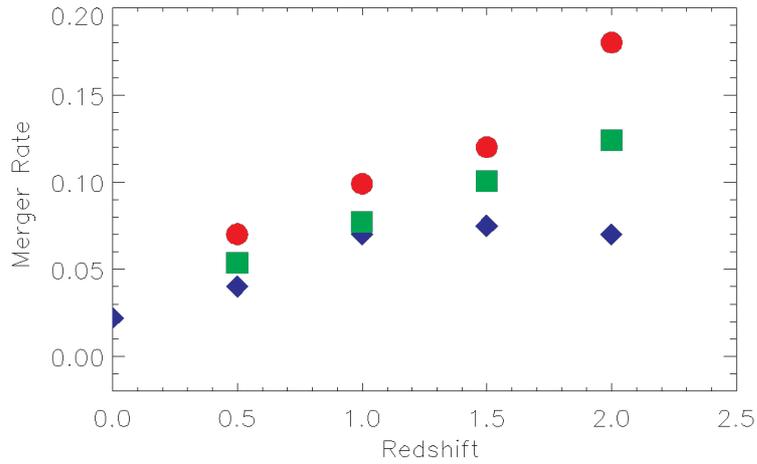

**Figure 6.** Observed (red circles) and simulated (blue diamonds) merger rates for the most massive mergers above 100 billion solar masses from [25] For simplicity, we have produced an average (green squares) of the two functions. Based on this we obtain the retrograde fraction as a function of redshift displayed in Figure 5. We could have used the red circles or the blue diamonds and our conclusions would remain qualitatively the same.

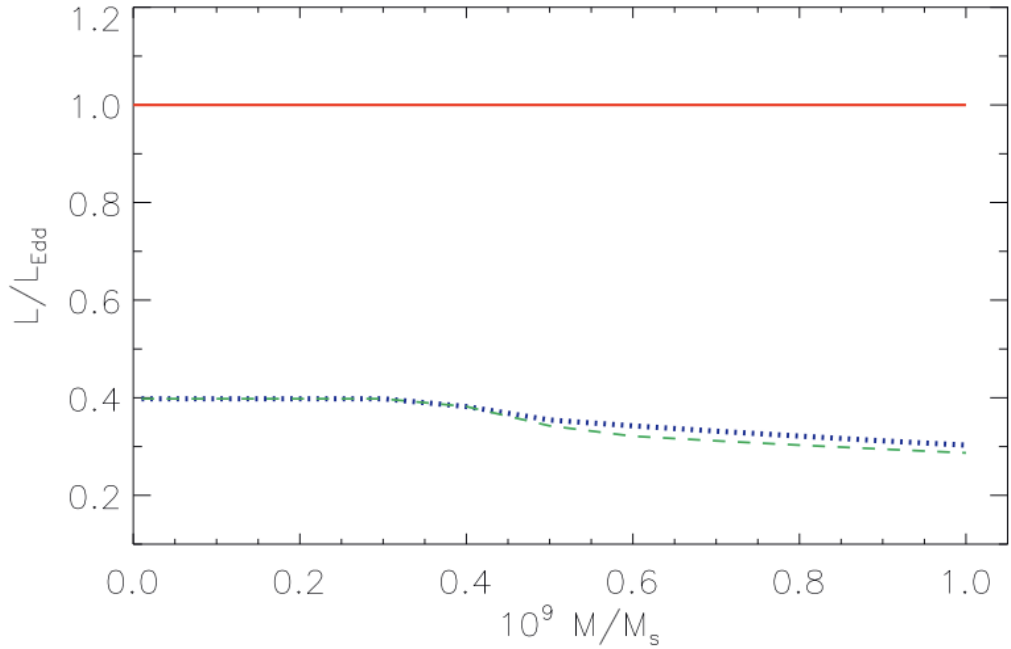

**Figure 7.** Average Eddington luminosity as a function of black hole mass (in billions of solar masses) from theory. The red line represents the Eddington limit while the blue and green represent the behavior at redshift below 1 and 1.5, respectively. Note the turnoff at higher black hole mass (above ≈ $4 \times 10^8$ M$_\odot$) resulting from the greater presence of less efficient Shakura & Sunyaev disks in retrograde configurations.

**Table 2.** Data for Figure 7. The left column is mass in solar masses. The right column is the Eddington luminosity or efficiency. Blue and green refer to redshifts 0.8 and 1.5, respectively. Because we have allowed a slightly larger fraction of retrograde black holes in the green data, the efficiency drops below that for the blue data.

| Mass (In Solar Masses) | L/L$_{Edd}$ (Blue) | L/L$_{Edd}$ (Green) |
|---|---|---|
| $10^7$ | 0.398 | 0.398 |
| $2 \times 10^7$ | 0.398 | 0.398 |
| $3 \times 10^7$ | 0.398 | 0.398 |
| $4 \times 10^7$ | 0.398 | 0.398 |
| $5 \times 10^7$ | 0.398 | 0.398 |
| $6 \times 10^7$ | 0.398 | 0.398 |
| $7 \times 10^7$ | 0.398 | 0.398 |
| $8 \times 10^7$ | 0.398 | 0.398 |
| $9 \times 10^7$ | 0.398 | 0.398 |
| $10^8$ | 0.398 | 0.398 |
| $2 \times 10^8$ | 0.398 | 0.398 |
| $3 \times 10^8$ | 0.398 | 0.398 |
| 4 × 10$^8$ | 0.382 | 0.382 |
| 5 × 10$^8$ | 0.354 | 0.342 |
| 6 × 10$^8$ | 0.342 | 0.321 |
| 7 × 10$^8$ | 0.331 | 0.312 |
| 8 × 10$^8$ | 0.321 | 0.303 |
| 9 × 10$^8$ | 0.312 | 0.295 |
| 10$^9$ | 0.303 | 0.287 |

In greater detail, for black hole masses ranging from 10$^6$ M$_\odot$ to about 10$^8$ M$_\odot$, none of the accreting black holes are allowed to accrete in retrograde configurations (i.e., Figure 5). Therefore, the quasar population at this black hole mass is restricted to prograde accreting black holes, which means they span a narrower range in ISCO values that are no larger than 6 GM/c$^2$. Furthermore, we assume that for a large range in spin magnitude, no black hole spin values are preferred or overrepresented with respect to others, so a straightforward average over the range of efficiencies for prograde accretion is determined. It is important to note that our constraints are not arbitrary as we are only imposing constraints on the retrograde versus prograde fraction and not on the individual spins which are determined by the physics of the merging black holes. This constraint ensures that average efficiencies must be above the efficiency for zero black hole spin. There are no free parameters that we can manipulate to obtain an average efficiency that is below that of a zero-spin accreting black hole. This constraint ensures that average efficiency at all redshifts be equal to an efficiency value corresponding to the efficiency of some intermediate spin prograde disk because no physical processes allow the retrograde regime to be over-represented at any time. In fact, it is precisely the opposite, with the prograde regime over-represented in the population. This analysis, therefore, allows us to produce theoretical points in the mass–luminosity plane. As we consider larger black hole masses, the data in Figure 5 indicates to us that a progressively greater range of ISCO values, and thus disk efficiencies, are included as a result of the increased presence of retrograde systems. Around 4 × 10$^8$ M$_\odot$, the fraction of retrograde occurrences is no longer negligible, which means that to obtain a data point in our theoretical plot for this mass, we must average over most of the range from retrograde to prograde efficiencies under the constraint that 0.125 of all the objects are accreting in the retrograde mode (Figure 5). As done for the prograde-only case, we do not limit nor constrain black hole spin (except as previously discussed at very high spin as mandated by numerical work), which means a wide range in spin values from high to low are not only represented, but equally so. This then produces an average efficiency for the 4 × 10$^8$ M$_\odot$ bin and another point on the mass–luminosity plane. The same analysis continues up to the highest black hole mass, where the average is over an equally represented range of both prograde and retrograde systems. It should be clear, therefore, that the average efficiency is decreasing with the increase in black hole mass.

To summarize the results, our explanation for the sub-Eddington boundary is that the most massive black holes produce less restrictive constraints on their accretion configurations, which determines, therefore, an average efficiency that is less than the average efficiency of less massive black holes that do, instead, place constraints on the formation of retrograde systems. While Figure 7 emerges from the precise prescription we use for the retrograde fraction as a function of mass, the uncertainties in this function can only result in a shift in the green line either toward the Eddington limit or further removed from it so the qualitative conclusion remains.

*2.2. IIb-Slope vs. Redshift of the Sub-Eddington Boundary*

In addition to the drop-off at higher black hole mass, [2] evaluate the slope of the fit at higher black hole mass as a function of redshift and find the slope to increase initially with redshift, peak, and subsequently decrease with higher redshift. They report the slopes in what is our Figure 8. We will incorporate this feature in our explanation. First, however, we want to observationally explore the [3] data as a function of redshift. We produce the equivalent of Figure 2 but at z > 4 in Figure 9. By looking at the observational data, one can identify whether a sub-Eddington boundary appears. We have explored the [3] data where the existence of a high mass sub-Eddington boundary is visible at the high mass limit around 10 billion solar masses, with the luminosities there falling shy of the $10^{48}$ erg/s of the Eddington limit (Figure 2). But we now can compare that to the data as a function of redshift. If we focus on the z > 4 data, the high mass sub-Eddington boundary is even more visible as a result of a flat distribution across the mass scale (Figure 9).

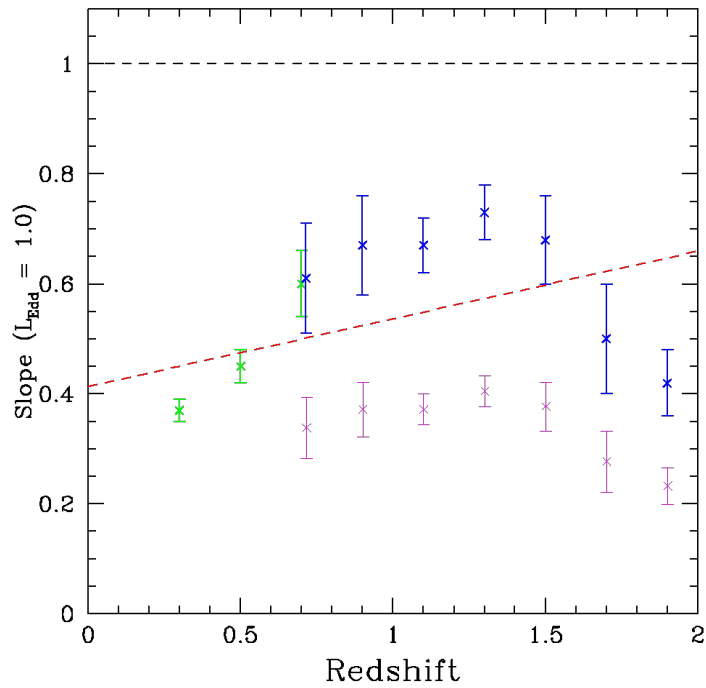

**Figure 8.** Redshift dependence of the slope of the Eddington boundary from Hβ (green), MgII (blue), and corrected MgII values (magenta). From [2] (Figure 9).

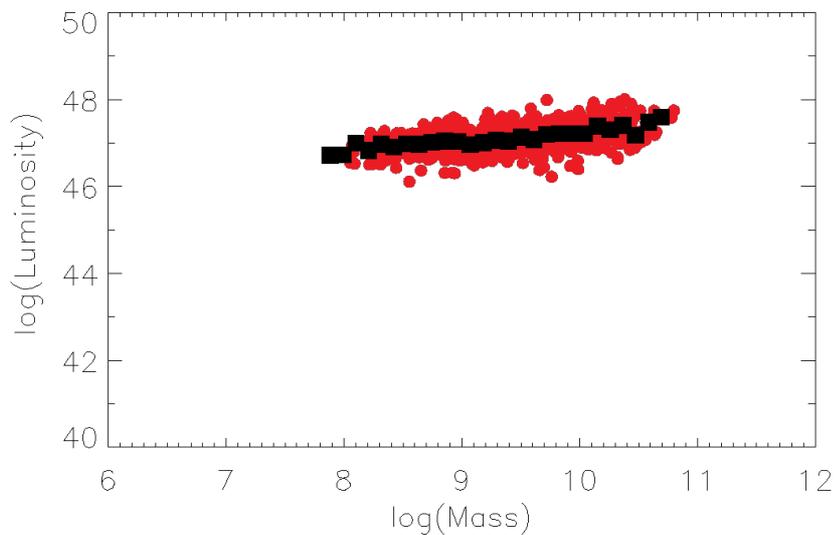

**Figure 9.** Luminosity versus Mass for redshift z > 4 from SDSS DR7 ([3]) with the average luminosity over-plotted as the black points.

In order to construct a redshift dependence in the slope, we must determine from theory the redshift dependence of the average luminosity at high black hole mass, which in turn requires that we evaluate how retrograde black hole formation depends on time. This, as we have already explored in Section IIa, forces us to consider the time dependence of the merger function (e.g., [25]as shown in Figure 6. Because the merger function peaks at z~1.5–2, this is the redshift corresponding to the greatest fraction of retrograde accreting black holes. In addition to the mass dependence of retrograde accretion, there is a time dependence as well that is directly connected to the time dependence of the physical mechanism that leads to retrograde accretion formation, namely galaxy mergers. As the redshift drops from z~1.5–2, the merger function drops and the fraction of retrograde accreting black holes drops in response to that. This connection between the fraction of retrograde accreting black holes and redshift was used to explain the observations of the shift in radio versus X-ray/optical peak in [26] (Figure 12), as well as the compatibility with the Soltan argument in [11] where the branching-tree diagrams in [11] show how the FRI LERGs described in this paper's Figures 10 and 11 are connected to the most massive black holes that began in retrograde mode. Here, we appeal to the same physics to evaluate the average luminosity as a function of redshift and to explain the time dependence of the slope on the mass–luminosity plane. Because the fraction of retrograde accreting black holes is largest, and thus the average efficiency is smallest, at z~1.5–2, this is also a time when the average luminosity struggles to reach Eddington values. According to Steinhardt & Elvis (2010), the slope of the sub-Eddington boundary at z~2 ranges from 0.2–0.4 depending on how it is measured (Figure 8). The slope at z~1, on the other hand, ranges from 0.65–0.35 for the blue and magenta points, respectively. In other words, the slope of the Eddington boundary at z~2 is about 60% of its value at z = 1 for both measurements. We construct the slope at both z~1.5 and z~0.8 by referring to the data in Table 2, but since we wish to explore the slope for the high mass population, we include Table 2 data only for black hole masses above $10^8$ solar masses. We include this data in the last two columns of Table 3. We plot these values in Figure 12.

In addition to the behavior of the slope of the sub-Eddington boundary at higher redshift (i.e., between z~1 and z~2), [2] also find a drop at redshifts below z~0.6. In other words, they find a peak in the slope of the sub-Eddington boundary for the highest black hole mass population at z~1, implying a decrease not only at higher z but also at lower z. At first glance, there seems to be tension between this observational fact and our explanation. Since the merger function continues to drop, in fact, the fraction of retrograde-forming objects continues to decrease. From the further drop postulated in our theoretical framework for the retrograde fraction as redshift decreases, we would expect the slope of the sub-Eddington boundary to continue increasing below z~1. However, this ignores the longer-term time evolution of the originally retrograde accreting objects that begins to have a dominant effect (Figures 10 and 11) by shifting the average to much lower disk efficiencies, as we now describe. The time evolution of powerful radio quasars involves a transition not only in black hole spin as the disk evolves into a prograde configuration from the originally retrograde one, but also in the type of accretion, which sees the radiatively efficient disk evolve into a hot, advection dominated (ADAF) flow on timescales of $10^7$ years or less ([27]; [28]; [7]; [11]). In other words, while the slope of the sub-Eddington boundary drops at higher redshift due to the lower efficiency of the average accreting high-mass black hole, the time evolution at lower redshifts of the efficiency involves competing effects. To appreciate the nature of this competition, see Figures 10 and 11. On the one hand, the ISCO decreases because continued accretion turns retrograde objects into prograde ones, which means that as long as the disk is radiatively efficient and thin, the efficiency increases. However, on a slower timescale, the state of accretion is transitioning toward advection-dominated, which means that despite the initial increase, the efficiency eventually not only drops but does so dramatically. This is true of all the Fanaroff-Riley II ([29]—FRII) quasars as can be seen in Figures 10 and 11. The ones with lower jet powers, in fact, also transition from a radiatively efficient thin disk to an advection-dominated disk. However, this takes longer than for the objects described in Figure 10. At the Eddington rate, the objects of Figure 10 approach the higher prograde regime after about 5 × $10^7$–$10^8$ years and eventually transition to lower efficiency ADAFs. Therefore, for the fraction of the most massive black holes that find themselves in retrograde accreting states, the efficiency of accretion begins low and starts to increase as ISCO values drop. But on the order of a few tens of millions of years, their efficiencies not only drop again, but do so dramatically. We argue this as the explanation for the data observed in Figure 8 and produce our own estimate of the slope of the Eddington boundary at lower redshift based on the above explanation. The objects we are referring to (FRI LERGs on the diagrams of Figures 10 and 11) are the massive, FRI radio galaxies, that accrete in low luminosity systems and that are thought to be responsible for the quenching of star formation [30].

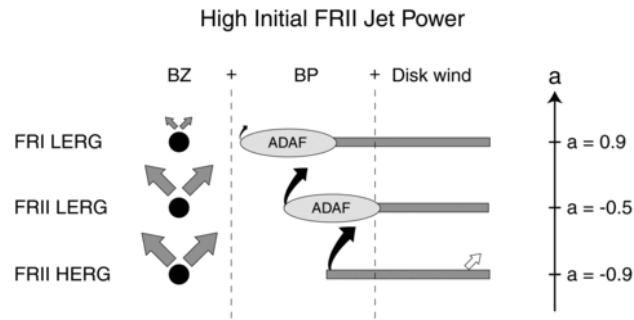

**Figure 10.** Time evolution of an initially retrograde accreting black hole with powerful jets (from [7]; where LERG is "low-excitation radio galaxies" and HERG is "high-excitation radio galaxies"). The radiative efficiency of the initially radiatively efficient thin disk (lower panel) evolves quickly into an advection-dominated disk, which has a radiative efficiency that is at least two orders of magnitude smaller than the Eddington value. BZ refers to the Blandford-Znajek jet ([31] and BP to the Blandford-Payne jet ([32]).

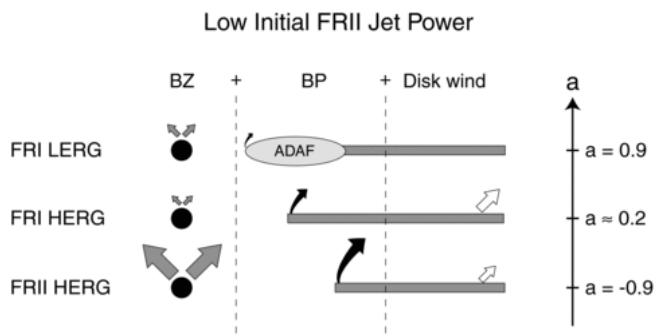

**Figure 11.** Time evolution of an initially retrograde accreting black hole with less powerful jets (from [7]; details same as Figure 10). The radiative efficiency of the initially radiatively-efficient thin disk (lower two panels) evolves less quickly (compared to the object in Figure 10) into an advection-dominated disk, which has a radiative efficiency that is at least two orders of magnitude smaller than the Eddington value.

The objects that remain radiatively efficient are generally part of the less massive of the massive black holes so they are less represented in this group. It is also worth noting that unlike in the discussion of [2] the drop-off at higher mass derived here is not a turn-off associated with cosmic downsizing, quite the contrary. It represents the early stage of evolution for some of the massive systems that accrete in retrograde mode. We have calculated the redshift dependence of the slope of the sub-Eddington boundary for three redshift values and report them in Figure 12. In comparing our results to those of [2] it is necessary to appreciate that for the slopes at redshift of 0.8 and 1.5 our results involve averaging only over the radiatively efficient class of accretors, and will therefore lead to a higher slope compared to [2] who, instead, deal with a distribution of objects that vary in Eddington luminosity by an order of magnitude. Our theoretical calculation is limited to accretion rates near the Eddington limit. Therefore, our theoretical slope must be larger by construction. For the data point at z~0.2, on the other hand, we explicitly include efficiencies that are much closer to the ADAF/thin disk efficiency at 0.01 Eddington. While FRII quasars are forming at z~0.2 at 1/4 the rate compared to z~1.5 (Figure 6), the distribution of accretion among the most massive black holes at z~0.2 is strongly influenced by the time evolution of their progenitors that have undergone the evolution described in Figures 10 and 11. This evolution involves both a transition toward the prograde accretion regime (which is what initially increases the efficiency and explains the data point at z~0.8) and a transition from radiatively efficient thin disk accretion into hot, advection-dominated, low-efficiency accretion. The efficiency for such objects spans a large range (orders of magnitude), but we are not interested in the lower or lowest efficiency ADAFs. We are still focused on objects that radiate efficiently or as efficiently as possible, so as to affect the Eddington boundary. Whereas at z~1.5 when thin disk accretion dominated the most massive black hole population and we averaged over the range of thin disk efficiencies above 0.1 Eddington as a result, here we also average over the most efficient accretors, but need to take into account the time evolution of the most massive black holes that were efficiently accreting in the past but are now evolving away from thin disk accretion. The effect of this time evolution for z~0.2 is the presence of thin, radiatively efficient accretion, that spans a wider range of efficiencies because the accreted material from the larger galaxy is hotter on average for the most massive black hole population compared to larger z. This forces us to average over all efficiencies that are associated with radiatively efficient accretion, namely accretion at or above 0.01 Eddington which is the theoretical boundary between ADAFs and thin disks. In order to account for the entire population of massive black holes accreting in radiatively efficient mode, we need to include the subgroup of AGN that were never strong jet producers and did not evolve from radio-loud quasar progenitors. A good fraction of the massive black hole population (approximately 1–0.43 and 1–0.35 at z~1.5–2 and z~1, respectively, as seen in Figure 5) were prograde accreting supermassive black holes and therefore highly-efficient radio-quiet quasars that, as long as they continued to be fed, remained prograde accreting systems. Hence, the distribution at z~0.2 is over a range of efficiencies from 0.8–0.01 to account for the radio-quiet quasar class. We produce an estimate that ranges from about 0.5 near the lower end of the massive black hole population at $3 \times 10^8$ solar masses to about 0.2 at the higher end near $10^9$ solar masses. The more massive black holes, in fact, are more likely to come from progenitor FRII quasars that were more effective in heating the galactic and intergalactic medium leading to the ADAF phase, and we try to capture that effect by lowering the efficiency slightly with an increase in black hole mass. Hence, the average efficiency for thin disk accretion at z~0.2 drops as a result of the presence of a large range of efficiencies that are close(r) to the efficiency that serves as the boundary between thin disk and ADAF. This explains the lowest point in Figure 10. While ours is an exact calculation given the postulated retrograde fraction in Figure 4, a change in that prescription will necessarily affect the z~0.2 point in Figure 12, further lowering that data point if we allow the fraction of retrograde objects to increase at lower black hole mass. Given this, the takeaway message should be that Figure 12 is qualitatively compatible with Figure 8.

**Table 3.** Log of $L/L_E$ vs. Log of black hole mass in units of 1 billion solar masses for z = 1.5, z = 0.8, and z = 0.2. The data for z = 0.8 and z = 1.5 was available in Table 2. However, we only used the data for black hole mass above $10^8$ solar masses.

| Log $M_9$ | Log ($L/L_{Edd}$) Z = 0.2 | Log ($L/L_{Edd}$) Z = 0.8 | Log ($L/L_{Edd}$) Z = 1.5 |
|---|---|---|---|
| 8.301 | 45.6 | 45.9 | 45.9 |
| 8.477 | 45.8 | 46.1 | 46.1 |
| 8.602 | 45.9 | 46.2 | 46.2 |
| 8.699 | 45.9 | 46.2 | 46.2 |
| 8.778 | 45.9 | 46.3 | 46.3 |
| 8.845 | 46.0 | 46.4 | 46.3 |
| 8.903 | 45.9 | 46.4 | 46.4 |
| 8.954 | 45.7 | 46.4 | 46.4 |
| 9.00 | 45.8 | 46.5 | 46.5 |
| Slope | 0.24 | 0.81 | 0.77 |

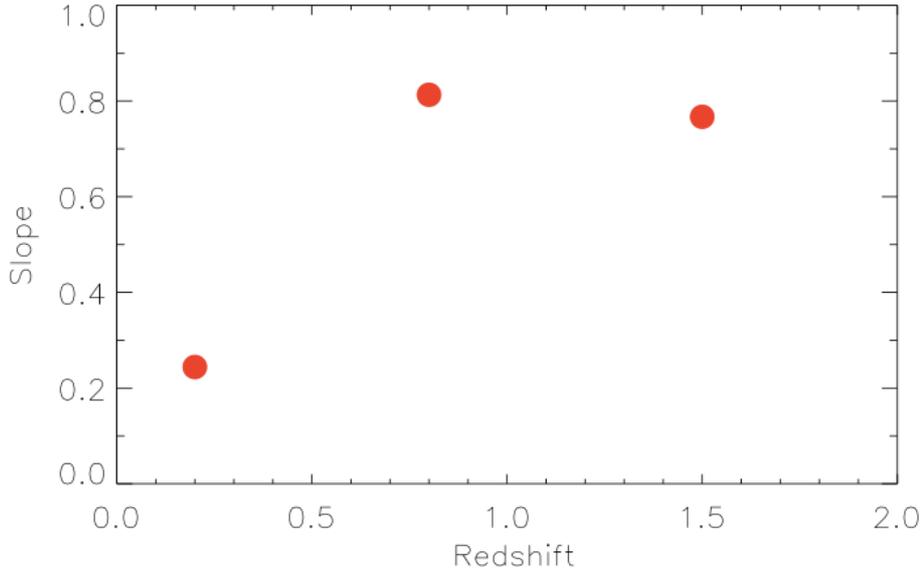

**Figure 12.** Redshift dependence of the slope of the Eddington boundary from theory. While the redshift 1.5 data point drops with respect to the redshift 0.8 point due to the larger fraction of radiatively efficient thin disks in retrograde configurations, the 0.2 redshift point drops due to the evolution of efficient accretors into advection dominated flows (ADAFs).

In closing, we point out that [5] have explored the $L/L_{Edd}$ ratios of ~58,000 type 1 quasars from the SDSS Data Release 7 and fail to find a sub-Eddington boundary, and [33] argues there is no sub-Eddington boundary in the [4] sample of ~28,000 SDSS quasars. At both z~3.2 and z~0.6, [5] find the $L/L_{Edd}$ distribution to be relatively independent of black hole mass. However, at redshifts $0.8 < z < 2.65$ the distribution of $L/L_{Edd}$ at fixed black hole mass shifts to larger Eddington ratios from black hole masses of ~5 × $10^8$ $M_\odot$ to $M_{BH}$ ~5 × $10^9$ $M_\odot$. This therefore implies that at $0.8 < z < 2.65$ type 1 quasars with more massive black holes are more likely to be radiating near the Eddington limit. It is important to emphasize that our conclusions are based on the presence of retrograde accreting black holes, which the model associates with broad line radio galaxies and FRII quasars. Absence of such objects in the population of radiatively efficient accretors will indeed increase the average $L/L_{Edd}$. We therefore caution observers to pay particular attention to an important difference for high black hole mass accretors from the perspective of the theoretical framework. If the higher black hole mass sample involves objects that have enough FRII quasars, the theoretical framework predicts that a sub-Eddington boundary will show up at higher mass. From the perspective of the paradigm, observers should verify whether the drop off on the mass luminosity plane exists for higher black hole mass quasars that involve FRII quasars, not simply for higher black hole mass quasars. If, in fact, the sample is primarily characterized by massive black holes in radio-quiet or non-jetted quasars, the paradigm prescribes the absence of a sub-Eddington boundary. A practical question addressing this issue, therefore, would be to ask if the fraction of radio loud or jetted quasars to the total quasar data is smaller in any subset of the SDSS data release 7 that is chosen for analysis compared to any subset of the SDSS data release 5. Because this work attempts to explain objects near the Eddington boundary— i.e., objects that are most luminous for a given black hole mass —Malmquist bias does not affect our results. It should be clear, i.e., that our analysis does not shed light on the range of luminosities per black hole mass but deals only with the largest luminosities at or near the Eddington boundary. We also emphasize that ours is not an attempt to strengthen the observational support in favor of a sub-Eddington boundary. We are, instead, focused on showing how such features emerge in a straightforward way from theoretical ideas that have been developed to explain a host of observations that are not directly related to the sub-Eddington boundary. In turn, even if it turns out that the observed sub-Eddington boundary is not physical, the theoretical ideas presented still incorporate it. It is important to point out that the ideas behind the sub-Eddington boundary as illustrated in Figures 10 and 11 are anchored to a prescription that makes many predictions but that fundamentally is grounded in the time evolution from retrograde to prograde accretion. In terms of jet morphology and excitation class, it is interesting to note that recent observations indeed fit within this prescription [34]. And, finally, in Figure 13 we show the number of quasars as a function of redshift, illustrating how a sub-Eddington boundary at a redshift of about 1.5 is even more significant than if the distribution were flat in the sense that more objects are available to reach their Eddington limit.

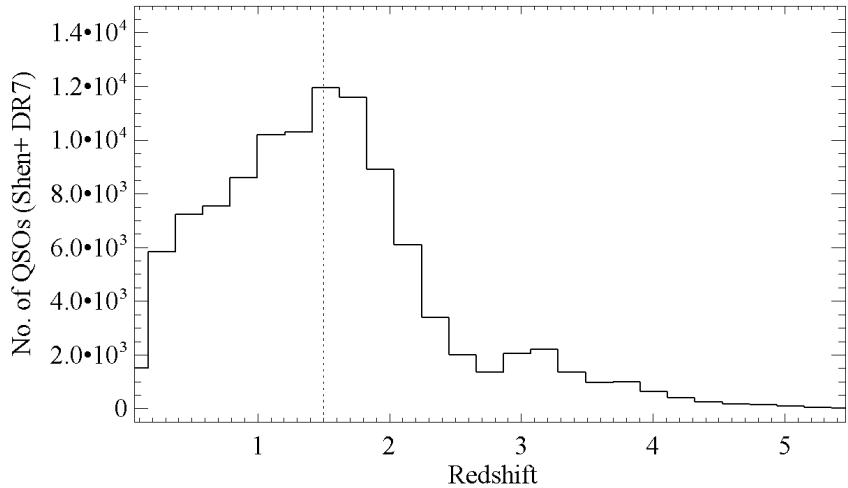

**Figure 13.** Distribution of quasars as a function of redshift from SDSS DR7 ([3]). Z = 1.5 is indicated with the vertical dashed line.

## 3. Conclusions

In this paper, we have attempted to unify various features of the sub Eddington boundary discovered by [2] within the context of the gap paradigm for black hole accretion and jet formation, a phenomenological framework anchored to the notion that retrograde accretion had an important role in the history of the most massive black hole mass buildup. While we have not produced any explanation as to why the Eddington boundary should be respected (which the observations, especially at lower black hole mass, do not require), we have shown how the range in Eddington ratio tends to vary as a function of black hole mass. We have shown, in other words, that if the basics of standard thin disk theory apply, the disk efficiency, which depends on black hole spin, must also depend on black hole mass. And the reason has to do with the post-merger fraction of retrograde black hole formation from which we find larger average ISCO values for the most massive accreting black holes. In addition to this efficiency versus mass dependence, we have shown that a redshift dependence emerges from the basic theory for the slope of the sub-Eddington boundary. Everything we have addressed in this paper comes from the connection between mergers and their ability to produce retrograde accreting thin disks around spinning black holes. While a number of assumptions are made for simplicity, they do not have an effect on the basic qualitative conclusion concerning the sub-Eddington boundary: Accreting black holes in retrograde configurations are less efficient than prograde ones which will contribute to lowering the average efficiency of the population. We have also pointed out that if observations are to constrain theory, an important distinction must be made between broad-line radio galaxies and FRII quasars and what are generally referred to as type 1 or 2 quasars.

**Author Contributions:** D.G. came up with the idea and wrote most of the paper. D.J.C. made many of the plots and contributed to the ideas as well as adding an analysis of the SDSS DR7 data. A.M.J. performed some of the calculations for his research project.

**Funding:** This research received no external funding.

**Acknowledgments:** We are grateful for the impressive and thorough review by the referees.

**Conflicts of Interest:** The authors declare no conflict of interest.